\documentclass{appolb}
\usepackage{epsfig}
\usepackage{graphicx}
\renewcommand{\thevolume}{??}
\renewcommand{\theyear}{2004}
\renewcommand{\theNo}{?}

\headtitle{Neutrinos from pre-supernova star}
\headauthor{A. Odrzywo\l{}ek et al.}
\begin{document}
\title{Neutrinos from pre-supernova star\thanks{
Talk presented at the Epiphany Conference on Astroparticle Physics,
8-11 January 2004, Cracow, Poland.}}
\author{
A. Odrzywo\l{}ek$^a$
\thanks{
{\tt odrzywolek@th.if.uj.edu.pl}
}\\
M. Misiaszek$^a$\\
M. Kutschera$^{a,b}$\\
\address{$~^{a}$M. Smoluchowski Institute of Physics, Jagiellonian University,
Krak\'{o}w,
Poland\\
$~^{b}$H. Niewodnicza\'{n}ski Institute of Nuclear Physics,
Krak\'{o}w,
Poland
} }
\maketitle
{\it  (Received April 30, 2004)}

\begin{abstract}
Analysis of the massive star properties during
C, Ne, O \& Si burning \ie the neutrino-cooled stage,
leads to the simplified neutrino emission model.
In the framework of this model we have simulated
spectrum of the antineutrinos. Flux normalized
according to the massive star model with explicitly
given neutrino luminosity allow us
to predict signal produced in water Cherenkov detectors.
The results are discussed from the point of view of the possibility of the
{\sl core-collapse} supernova event prediction in advance of a few days.
\end{abstract}

\section{Introduction}

In this paper, we will overview problems discussed in more detail in our article {\em ``Detection possibility of the pair-annihilation neutrinos from neutrino-cooled pre-supernova star''} \cite{astroparticle}. The conference presentation can be viewed from the Epiphany 2004 homepage \cite{presentation}.

Until now, one has detected neutrinos from two specific astrophysical sources:
the Sun and supernova SN~1987A.
Solar neutrinos were detected because of proximity, and steady,
uninterrupted emission. They were first detected by Raymond Davis
chlorine detector at Homestake.
Neutrinos from supernova 1987A, located far away, in our neighbor galaxy, The Large Magellanic Cloud,
were detected because of extremely strong neutrino burst. This short burst that lasted about 1 minute carried enormous energy of $10^{53}$ ergs released in the gravitational collapse of the stellar core.

There exist a number of the proposed astrophysical neutrino sources expected to be possible to detect in the future, \eg collapse of super-massive ($M>100 M_\odot$) stars \cite{fuller}. We would like to present here our proposition which is detection of the neutrino-cooled stars.

\section{Neutrino-cooled stars}

Neutrino-cooled star is a well-known astrophysical object, but under new, much more relevant name. This terminology was introduced in the excellent textbook of David Arnett \cite{arnett}.
Shortly speaking, neutrino-cooled star is almost the same object as massive star or pre-supernova star, but at different evolutionary stage.

By the definition, massive star, is a star which is able
to explode as the {\sl core-collapse} supernova.
Theory of the stellar evolution tell us
that initial mass of such star is about 8-9 M$_\odot$ or more.
We define:
\begin{itemize}
\item{Neutrino-cooled star -- the massive star after ignition of the carbon burning}
\item{Pre-supernova star -- the massive star at the onset of the collapse}
\end{itemize}
The word `{\it onset}' is not very precise. One defines frequently
the last model computed by the stellar evolution hydrostatic code as
a pre-supernova star. Usually, the sequence of models is terminated
if the inward velocity at some point (\eg at the edge of the `Fe' core)
exceeds some previously defined value of \eg $1000$~km/s \cite{heger3}.
Nevertheless, from the external observer point of view, it is not
a very big mistake, to refer to neutrino-cooled star as pre-supernova,
because the evolution after carbon ignition is very rapid. It takes only 300 years
for $20 M_\odot$ star to exhaust whole fuel. This time is very short compared
to both the entire lifetime of a star ($\sim$10 mln years) and the Kelvin-Helmholtz
timescale of hydrogen envelope ($\sim$10 000 years).

Introduction of  the neutrino-cooled stars is not only the matter
of enriched astrophysical terminology, but also reflects the important
physical differences between these stars and \eg the main sequence stars.
The massive stars after carbon ignition are completely different from other
stars.

Basic comparison of typical neutrino-cooled star, represented by the 20~M$_\odot$ model of Woosley \cite{woosley}, and the main sequence star, represented by the Sun, reveals some very important facts.
Lifetime of massive star is very short, about 10 million years,
compared to 10 billion years for the Sun.
Neutrino-cooled stage lasts only about 300 years.  This is relatively short period. That is why some of astrophysicists think about neutrino-cooled events rather than neutrino-cooled stars.

The most important difference between these two stars is the process of cooling.
Comparison of the photon and neutrino luminosities of our stars, lead to
the following conclusions.
Neutrinos are only small part ($\sim$2\%) of the Solar energy budget. The opposite situation
takes place in neutrino-cooled stars. The neutrino luminosity is impressively big, up to $10^{12} L_\odot$. The photon luminosity of $10^5 L_\odot$ is almost
negligible\footnote{This value is still a hundred thousands times
greater than solar luminosity $L_\odot$.}
compared to the neutrino luminosity.
The neutrino luminosity of $10^{12} L_\odot$ could explain the name: neutrino-cooled star, especially if we realize that this is $10^7$ times more than the photon luminosity.

To imagine how big the neutrino luminosity is, let us make some very rough estimates.
We will refer to our knowledge of
supernova and solar neutrinos.
Peak neutrino luminosity from {\sl core-collapse}
supernova reaches $2.5 \cdot 10^{53}$ erg/sec $\sim$10$^{20} L_\odot$, but it lasts only a few
milliseconds \cite{burrows}. The average neutrino luminosity during $\sim$100 seconds
of the main protoneutron star cooling phase is $\sim$10$^{51}$ erg/sec i.e. $\sim$10$^{17} L_\odot$.
This is still five orders of magnitude greater than the neutrino
luminosity of $\sim$10$^{12}$ L$_\odot$ during silicon burning. If we take into account
much longer time of silicon burning of a few days $\sim$10$^5$ seconds,
we realize that the total amount of pre-supernova neutrinos
may reach 1\% of the total energy released during the {\sl core-collapse} supernova event.
In other words, if we are able to detect supernova neutrinos from
`$x$' kiloparsecs, we could possibly detect Si-burning neutrinos from
distance of `$x/10$' kpc. Unfortunately, the energy of 10-20~MeV,
typical for supernova neutrinos, is highly unlikely
for pre-supernova neutrinos. Additionally, from the experimental point of view,
it is much more difficult to detect the same number of the neutrinos but emitted
in much longer time.

Solar neutrinos appear to have energy similar (on average) to
pre-supernova neutrinos \cite{astroparticle}.
We may then ask a simple question: {\it From what distance neutrino-cooled star will ``shine'' on the ``neutrino sky'' like the Sun?} Precisely, from what distance $D$
to the massive star, the neutrino flux on the Earth will be the same for both the neutrino-cooled star and for the Sun?
Very simple expression gives the required distance in astronomical units:
\begin{equation}
{D} = \sqrt{
\frac{\vphantom{\frac{\int}{1}}10^{12} \, \mathrm{L}_\odot}
{\vphantom{\frac{1}{\int}}0.02 \, \mathrm{L}_\odot}} =
6.2 \cdot 10^6 \mathrm{AU},
\label{dist}
\end{equation}
as only 0.02~L$_\odot$ of the energy released in the Sun's core is carried off by the neutrinos.
Distance (\ref{dist}) is equal to thirty parsecs or 100 light years.

One of the most close known stars similar to our neutrino-cooled star is Betelgeuse
($\beta\;Ori$), 185 parsecs away. This is five times further away than (\ref{dist}), so we may conclude that the detection of neutrinos from such star is impossible,
because simply no such close star exist!

Fortunately, this is a wrong conclusion, because the neutrino emission from the pre-supernova star is different compared to the solar neutrinos.

\section{Pair-annihilation neutrinos}

\subsection{Spectrum of the pair-annihilation neutrinos}

In the neutrino-cooled stars, the energy generation is balanced by the neutrino emission.
Energy sources are the nuclear reactions and gravitational energy.
They are balanced by the emission of the thermal and weak-nuclear neutrinos.
The former dominates, but the latter is progressively more intense and finally (during Si burning) becomes the most important \cite{langanke}.
A lot of processes can produce thermal neutrinos \cite{itoh}.
The most dominant process is annihilation of the electron-positron pairs into neutrinos. This process requires very high temperature of $\sim10^9$~K.
\begin{figure}
\includegraphics[width=\textwidth]{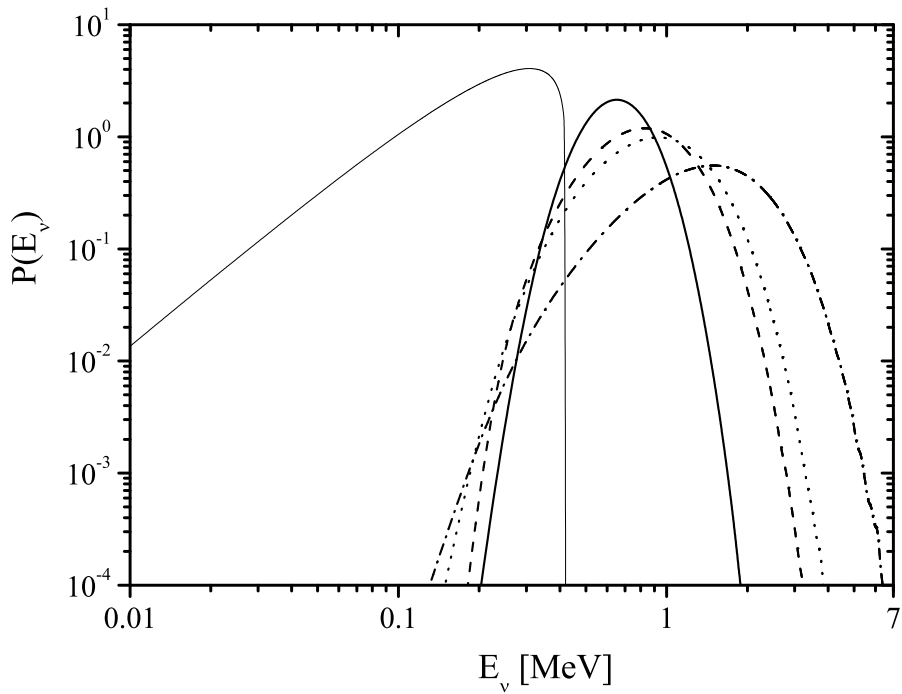}
\caption{Normalized spectrum of the pair-annihilation neutrinos ($\bar{\nu}_e$) emitted during
C (solid), Ne (dashed), O (dotted) and Si (dot-dashed) burning stage \cite{astroparticle} and the solar $pp$ neutrinos (thin line) \cite{bahcall}. Average $\bar{\nu}_e$ energy is $1.85$ MeV during Si burning, but spectrum extends
up to $\sim$6~MeV. The solar neutrino spectrum is known to be
more complicated. Possibly, the same holds for the pre-supernova star,
but this requires further research.  See also Fig.~1 of Ref.~\cite{astroparticle} or \cite{presentation} where the massive star is at distance of 1~kpc, while the Sun is still at distance of 1~AU.}
\label{solar_vs_presupernova}
\end{figure}

We have calculated spectrum of the pair-annihilation neutrinos \cite{astroparticle}. The result is presented in Fig.~\ref{solar_vs_presupernova} together with solar $pp$ neutrinos.
Actually, neutrino spectrum of the pre-supernova star is
possibly more complicated due to numerous nuclear and thermal processes. Little is known about this spectrum. Besides of our calculations of the pair-annihilation
process, only the spectrum of the six $e^{-}$ capture processes on the `Fe' nuclei during Si shell burning has been calculated \cite{langanke2}. Some of these neutrinos
have energy of $\sim$8~MeV. This indicates that detailed study
of neutrino-cooled star neutrino spectrum may reveal ``solar $^8$B $\nu_e$ equivalent'',
leading to much easier and well-understood detection techniques
compared to those discussed in this article.

As we can see (Fig.~\ref{solar_vs_presupernova}), the spectrum is different, and in some energy range, close to 1-2~MeV, the detection of the neutrino-cooled star may be easier than the detection of the solar $pp$ neutrinos.
But the most important difference, not visible in Fig.~\ref{solar_vs_presupernova}, is the presence of antineutrinos.
Pair-annihilation neutrinos are produced in the reaction:
\begin{equation}
e^{+} + e^{-} \longrightarrow \nu_x + \bar{\nu}_x,
\end{equation}
where $x = e, \mu, \tau$. About $1/3$ of the flux is
in the form of the electron antineutrinos \cite{astroparticle}.
In contrast, the Sun does not emit antineutrinos of any flavor at all.

\subsection{Detection using inverse $\beta$ decay}

As we have pointed out (\cite{astroparticle}, Sect.~5) the presence of the antineutrinos
is very important, because we may use very efficient reaction of inverse beta-decay for the detection purpose:
\begin{equation}
\bar{\nu}_e + p \longrightarrow n + e^{+}.
\label{inv_beta}
\end{equation}
This reaction for the Si burning neutrinos has large spectrum-averaged
cross-section
$\bar{\sigma}_{\mathrm{Si}} \simeq 0.7 \cdot 10^{-43}\, \mathrm{cm}^2$ \cite{astroparticle}.
This is 3 times more than \eg the cross-section for elastic
neutrino-electron scattering.

\subsection{Event rates in H$_2$O}

We have estimated event rate, which is
the number of reactions (\ref{inv_beta}) induced by the pair-annihilation neutrinos in the water.
Event rate $r$ is:
\begin{equation}
r \, [day^{-1}] = f\,\cdot \bar{\sigma}_\alpha \, [cm^2] \cdot
N \cdot \phi_\alpha \, [ cm^{-2} \, day^{-1}],
\end{equation}
where $f$ is $\bar{\nu}_e$ fraction, $\bar{\sigma}$ is the averaged cross-section,
N is the  number of targets, $\phi$ is the flux on Earth; $\alpha \,=\,$C, Ne, O, Si refers to a burning phase.
During silicon burning, for 20~M$_\odot$  star, from distance of 1~kpc, in one kiloton of water, pair-annihilation neutrinos will produce 1.3 neutrons per day
\cite{astroparticle}.
This result is easy to remember: we have approximately one reaction per day in one kiloton of the water from distance of 1 klioparsec.
This is relatively high event rate, especially in the volumes of the future megaton-scale detectors: Hyper-Kamiokande \cite{hk} and  UNO \cite{uno}.

Let us summarize:
\begin{itemize}
\item{giant source of the electron antineutrinos has been identified}
\item{reaction with large cross-section for the detection purposes has been chosen}
\item{expected reaction (\ref{inv_beta}) rate in the water is relatively high}
\end{itemize}
The most confusing problem is how to detect
the products of the reaction (\ref{inv_beta}) in giant volumes of water detectors.

\subsection{Inverse $\beta$ decay in water detectors}

The detection of the reaction (\ref{inv_beta}) products in water is very difficult.
The positron energy $E_{e^{+}} \simeq E_{\nu}-1.8$~MeV is too low
to produce detectable Cherenkov light.
Neutrons are captured on protons, producing invisible 2.2 MeV gamma-ray
cascade. Nevertheless, chance for detection exists, as it is discussed in the next subsections.

\subsection{Detection of $e^{+}$}

Both $e^{+}$ and $e^{-}$ are detected in water detectors
via emission of Cherenkov light. In Super-Kamiokande detector,
reaction (\ref{inv_beta}) rate  for Si burning neutrinos from
star at distance of 1~kpc is 41/day. Because of the
threshold energy only a fraction of positrons produced
by (\ref{inv_beta}) will be detected\footnote{In contrast, supernova
$\bar{\nu}_e$'s produce $e^{+}$ well above detector threshold, with almost
100\% detection efficiency.}. This fraction
very strongly depends on the actual value of threshold (\cf~Table~\ref{events_vs_threshold}).
For single events threshold may be as low as $\sim4$ MeV in SK \cite{kielczewska}, but still only 0.7\% of reactions (\ref{inv_beta}) could be detected.
Without dramatic improvements of experimental technique
detection of positrons alone seems to be hopeless.

\begin{table}
\begin{center}
{\sc Table I}\\
\caption{Number of the detectable positrons for given detector
threshold in Super-Kamiokande. The neutrino energy $E_{\nu}$ required to
produce positron of energy $E_e$ (including rest mass) is greater than
$E_\nu = E_e + \Delta$,
where $\Delta=m_n-m_p$ is difference between neutron and proton mass.
Using coincidence with $n$ capture on Cl or Gd, reduction of the threshold
may be possible, as indicated by an arrow.}
\label{events_vs_threshold}
\fbox{
\begin{tabular}{c|c||cc}
$E_{th}^{\bar{\nu}_e}$ & $E_{th}^{e^{+}}$ & Detectable positrons &\\
{\large [MeV]} & {\large [MeV]} & (day)$^{-1}$&\\
\hline \hline
1.8	&0.5	&41&\\
3.0	&1.7	&22&\\
\hline
4.0	&2.7	&6.5& $\uparrow$\\
5.0	&3.7	&1.2& SK\\
6.0	&4.7	&0.2 &$|$\\
\hline
7.0	&5.7	&0.0&
\end{tabular}
}
\end{center}
\end{table}

\subsection{Detection of neutrons}

We are able to detect neutrons produced by reaction (\ref{inv_beta})
if we dissolve in water some
efficient neutron absorber. In SNO \cite{sno} neutrons are detected using capture
on Cl (dissolved salt) nucleus. Other proposed method is to dissolve GdCl$_3$ in water \cite{gadzooks}, with gadolinium being very effective neutron absorber.
Neutrons will be captured on these (Cl, Gd) nuclei:
\begin{equation}
n + \mathrm{Cl} (\mathrm{Gd}) \longrightarrow \mathrm{Cl}^{\ast} (\mathrm{Gd}^{\ast})
\longrightarrow \mathrm{Cl} (\mathrm{Gd}) + \gamma.
\end{equation}
The excited nucleus emits cascade of the gamma-rays with total energy of $\sim$8 MeV.
High energy photons hit  electrons, and scattered electrons (with energy above
detector threshold) radiate Cherenkov light. Cherenkov light is detected by the photomultipliers.

SNO experiment has proved that neutron detection
efficiency with use of this method is close to 100\% \cite{sno}. Nevertheless,
high background level (\eg $\sim$100 events/day for Super-Kamiokande)
allow us to detect neutrino signature of the massive star silicon burning
stage only for a few very close stars, \eg Betelgeuse \cite{snwarning}.

\subsection{Coincidence detection of both $e^{+}$ and $n$}

Overall detection efficiency for low ($\sim$few~MeV) energy antineutrinos
by reaction (\ref{inv_beta})  can possibly be greatly improved if
coincidence of neutron capture and $e^{+}$ Cherenkov light will be used.
Using coincidence, we are able to reduce detector threshold \cite{sno}.
This is extremely important (\cf~Table~\ref{events_vs_threshold}),
because number of detectable events rises very fast with reduced threshold
energy. Hopefully, values as low as $2.5$ MeV for recoil electron energy
could be reached \cite{gadzooks}. Such low value of the threshold, for silicon burning antineutrinos with spectrum
presented in Fig.~\ref{solar_vs_presupernova} allow us to
detect 10\% of reactions (\ref{inv_beta}) occurrences in the water Cherenkov detectors. These coincidence events could be distinguished from background signal
much easier than single neutrons.
 In scintillator detectors, \eg KamLAND, gammas from annihilation of the positron
produced in the reaction (\ref{inv_beta}) give additional
1~MeV of the detectable energy.

\section{Conclusions}

If we solve all problems related to the detection of neutrons and positrons, produced by antineutrinos in water Cherenkov or other detectors in the reaction (\ref{inv_beta}) new, amazing possibility will open.

In Super-Kamiokande detector, enhanced by dissolved neutron absorber (NaCl, GdCl$_3$) the signal produced by the neutrinos from Betelgeuse (15 M$_\odot$ red giant at distance 185pc)
during Si burning could be $\sim$1000 events/day \ie 10 times more than
current background level \cite{snwarning}. Therefore, prediction
of supernova explosion for nearby stars in advance of a few days is possible.
Unfortunately only a few such close stars
exist\footnote{One of them is binary system $\gamma^2$ Velorum
at distance of 258~pc, which consists of $\sim$9.5M$_\odot$ Wolf-Rayett star
and $\sim$30M$_\odot$ O star \cite{gamma-2}.}, with
extremely small explosion probability in, say, next 100 years.
Therefore, the question is from what distance we will be able
to detect Si burning neutrinos. In Super-Kamiokande,
detection of neutrons captured by the Cl or Gd is possible only if the number of events exceeds background signal of $\sim$100 events/day.
Such neutron signal results from 20~M$_\odot$ star at distance of $\sim$600~pc.
Much more promising is detection using coincidence
of $n$ capture with $e^{+}$. Assuming that every single such coincidence event/day with $e^{+}$ energy above $2.5$ MeV can be distinguished from the background, we
get the maximum observation range of 2~kpc. This is
close to gravitational radiation from {\sl core-collapse} supernova detection abilities
of LIGO~I detector of 5~kpc \cite{muller}.

If megaton-scale detectors, like proposed Hyper-Kamiokande (540 kilotons) \cite{hk} or UNO (440 kilotons) \cite{uno} do appear, we will be able to extend observational range significantly. We may expect, for Hyper-Kamiokande,
increase of detectable stars range by a factor of $\sqrt{540/32}\simeq4$ due to target volume
much larger than Super-Kamiokande (of only 32 kilotons). Up to a distance
of 8~kpc we expect to find 35\% of the Galaxy disk stars \cite{galaxy_model}.
For the closest massive stars (Betelgeuse, $\gamma^2\;Vel$), such giant detectors allow us even to attempt detection of Ne and O burning antineutrinos with
number of reactions (\ref{inv_beta}) of 2/day and 45/day, respectively. These
burning stages precede core collapse by a few months. This could
give us insight into pre-supernova star burning processes, and establish
initial conditions for {\sl core-collapse} supernova explosion.
It could also provide us with very early warning of subsequent
core collapse. This allow scientists to prepare all available
observational devices, being down \eg due to maintenance, especially gravitational radiation detectors. It could help not to miss the most expected
astrophysical event.

\end{document}